\documentclass[aps,pra,twocolumn,superscriptaddress,showpacs,nofootinbibfloatfix,amsmath,amsfonts,amssymb]{revtex4-2}%

\usepackage{graphicx}
\usepackage{float} 
\usepackage{subfigure}
\usepackage{amsmath}
\usepackage{amsfonts,amscd,amssymb,amsthm}
\usepackage{booktabs}  
\usepackage{verbatim}
\usepackage{bm}
\usepackage{amstext}

\usepackage{amsmath,amsfonts,amssymb,color}
\usepackage{amsthm}
\usepackage{leftidx}
\usepackage{graphicx}
\usepackage{xcolor}
\usepackage{dcolumn}
\usepackage{bm}
\usepackage{epstopdf}
\usepackage{epsfig}
\usepackage{environ}
\usepackage{pdfcomment}

\usepackage{multirow}
\usepackage{setspace}
\usepackage{color}

\usepackage{float}
\usepackage[T1]{fontenc}
\usepackage{setspace}
\usepackage{esint}

\begin{document}
	
\title{Two-body interaction induced phase transitions and intermediate phases in nonreciprocal non-Hermitian quasicrystals}
	
\author{Yalun Zhang}
\affiliation{%
College of Physics and Optoelectronic Engineering, Ocean University of China, Qingdao 266100, China
}

\author{Longwen Zhou}
\email{zhoulw13@u.nus.edu}
\affiliation{%
College of Physics and Optoelectronic Engineering, Ocean University of China, Qingdao 266100, China
}
\affiliation{Institute of Theoretical Physics, Chinese Academy of Sciences, Beijing 100190, China}
\affiliation{%
Key Laboratory of Optics and Optoelectronics, Qingdao 266100, China
}
\affiliation{%
Engineering Research Center of Advanced Marine Physical Instruments and Equipment of MOE, Qingdao 266100, China
}
 
\date{\today}
	
\begin{abstract} 
Non-Hermitian phenomena, such as exceptional points, non-Hermitian skin effects, and topologically nontrivial phases have attracted continued attention. In this work, we reveal how interactions and nonreciprocal hopping could collectively influence the behavior of two interacting bosons on quasiperiodic lattices. Focusing on the Bose-Hubbard model with Aubry-Andr\'e-Harper quasiperiodic modulations and hopping asymmetry, we discover that interactions could enlarge the localization transition point of the noninteracting system into an intermediate mobility edge phase, in which localized doublons formed by bosonic pairs can coexist with delocalized states. Under the open boundary condition, the bosonic doublons could further show non-Hermitian skin effects, realizing doublon condensation at the edges, and their direction of skin-localization can be flexibly tuned by the hopping parameters. A framework is developed to characterize the spectral, localization, and topological transitions accompanying these phenomena. Our work advances the understanding of localization and topological phases in non-Hermitian systems, particularly in relation to multiparticle interactions.
\end{abstract}

\maketitle
	
\section{Introduction}\label{sec.int}
Anderson localization (AL) refers to the absence of diffusion in
disordered media \cite{anderson1958absence,Mott1978RMP,Gof41979PRL}. As a generic phenomenon, it plays
an important role in both classical and quantum transport of waves under
spatial randomness~\cite{Lagendijk2009PhysTody}. In quantum regime, it also describes a
class of metal-insulator phase transition driven
by disorder \cite{Belitz1994RMP,Evers2008RMP}. Over the past six decades, the AL has attracted continued
interest in both theoretical \cite{anderson1980new,fishman1982chaos,lee1985disordered,punnoose2005metal,lemarie2010critical,mott2012electronic,abanin2019colloquium} and experimental~\cite{billy2008direct,Roati2008Nat,Deissler2010NatPhys,Schreiber2015Sci,choi2016exploring,meier2018observation} studies.
However, the fate of localization in open systems and under many-body interactions
is still a challenging issue that has not been fully resolved.

Non-Hermitian systems, usually characterized by non-Hermitian Hamiltonians
constitute ideal setups to explore the interplay among localization, environmental
effects and interactions. In recent years, non-Hermitian systems have
attracted increasing attention due to their distinctive physics, such
as the parity-time-reversal (PT) transitions~\cite{bender1998real,guo2009observation,peng2014parity}, exceptional
points~\cite{berry2004physics,heiss2012physics,zhen2015spawning,ding2016emergence,midya2018non,miri2019exceptional}, non-Hermitian skin effects (NHSEs) \cite{yao2018edge,kunst2018biorthogonal,martinez2018non,lee2019anatomy,okuma2020topological} and
enriched symmetry classifications of topological phases \cite{gong2018topological,shen2018topological,kawabata2019symmetry,liu2019topological,kawabata2019topological,wojcik2020homotopy,shiozaki2021symmetry}.
Meanwhile, the interplay between disorder and non-Hermitian effects
may lead to unique types of spectral, localization, topological and
entanglement transitions \cite{hatano1996localization,asatryan1996electromagnetic,feinberg1999non,jazaeri2001localization,basiri2014light,mejia2015interplay,hatano2016chebyshev,zeng2017anderson,jiang2019interplay,longhi2019topological,longhi2019metal,huang2020anderson,tzortzakakis2020non,zeng2020topological,kim2021disorder,claes2021skin,cai2021boundary,tang2021localization,liu2021localization,cai2021localization,longhi2021non,longhi2021phase,liu2022topological,sarkar2022interplay,lin2022observation,acharya2022localization,zhou2022topological,cai2022localization,chen2022quantum,orito2022unusual,orito2023entanglement,li2023disorder,zhou2024entanglement,li2024emergent}, yielding universality classes
of localization beyond closed-system contexts \cite{kawabata2021nonunitary,luo2021transfer,luo2021universality,chen2024field}.
With additional elements like lattice dimerization, long-range hopping,
non-Abelian potential and time-period driving, intermediate phases and
mobility edges could also emerge in disordered non-Hermitian systems
even in one spatial dimension \cite{zeng2020winding,liu2020non,liu2020generalized,zhou2021floquet,zhou2021non,wang2021unconventional,Xu2021PRB,liu2021exact,han2022dimerization,liu2021exact_nh,yuce2022coexistence,xia2022exact,longhi2022non,weidemann2022topological,lin2022topological,zhou2022driving,zhou2023non,Wang2024PRB}, enriching further the non-Hermitian
localization phenomenon.

Interaction could greatly complicate the AL in non-Hermitian
systems due to its active competition with other localization mechanisms,
including random or correlated disorder and NHSEs.
Great efforts have been made towards the understanding of ``many-body
localization'' in non-Hermitian settings, with a focus on multiple
interacting particles in lattices of relatively small sizes \cite{levi2016robustness,medvedyeva2016influence,hamazaki2019non,panda2020entanglement,zhai2020many,wang2021many,heussen2021extracting,suthar2022non,li2023non,ganguli2024spread,li2024imaginary}.
Nevertheless, considering a few interacting particles in relatively
large lattices may offer necessary insights regarding how
interactions could drive different phase transitions and criticality
in the thermodynamic limit (with system
size $L\rightarrow\infty$) \cite{zhang2013bound,di2016two,gorlach2017interaction,azcona2021doublons,mark2020interplay,Longhi2024PRB,qian2024correlation,rassaert2024emerging,liu2024fate}. Motivated by these considerations,
we explore in this work the interaction-driven phase transitions and
intermediate mobility edge phases in non-Hermitian quasicrystals (NHQCs). Focusing on
two interacting bosons in one-dimensional (1D) Aubry-Andr\'e-Harper
(AAH) lattices with quasiperiodic onsite potentials and
nonreciprocal hoppings, we reveal that the interactions could not only
transform the AL critical points of noninteracting systems into intermediate mobility edge
phases with coexisting extended and localized states, but also push the boundaries of fully localized phases towards
stronger asymmetric hopping regions. Within an intermediate phase, extended
two-boson states can coexist with bounded boson pairs forming doublons,
which could also subject to NHSEs under the open boundary condition (OBC).
The borders between extended, intermediate and localized phases are further
found to carry topologically nontrivial signatures.

The rest of the paper is organized as follows. In Sec.~\ref{sec.model}, we introduce the model considered in this work,
summarizing its known features in the noninteracting limit, and presenting the methods we take to study its physical properties. In Sec.~\ref{sec.result}, we
characterize the real-complex spectra transitions, localization transitions, topological
transitions and intermediate mobility edge phases in representative NHQCs with
asymmetric hoppings, tracing their origins
back to the interplay between interactions and non-Hermitian effects of two interacting bosons.
In Sec.~\ref{sec.doublon}, we focus on the strong-interaction regime and explore
the competition of two different localization mechanisms (disorder
vs NHSEs) on bosonic doublons. In Sec.~\ref{sec.conclusion}, we summarize our results
and discuss potential future directions.
	
\section{Model and methods}\label{sec.model}
This section introduces the NHQC model that we will consider in this work. Without interactions, features of the model regarding its spectral, localization and topological transitions will be recapped. With onsite Bose-Hubbard interactions, we introduce quantities to characterize the phases and transitions in our system, including the ratio of complex energies, inverse participation ratios (IPRs) and topological winding numbers.

\subsection{Model}\label{sec.GM}

\begin{figure}
\begin{centering}
\includegraphics[scale=0.38]{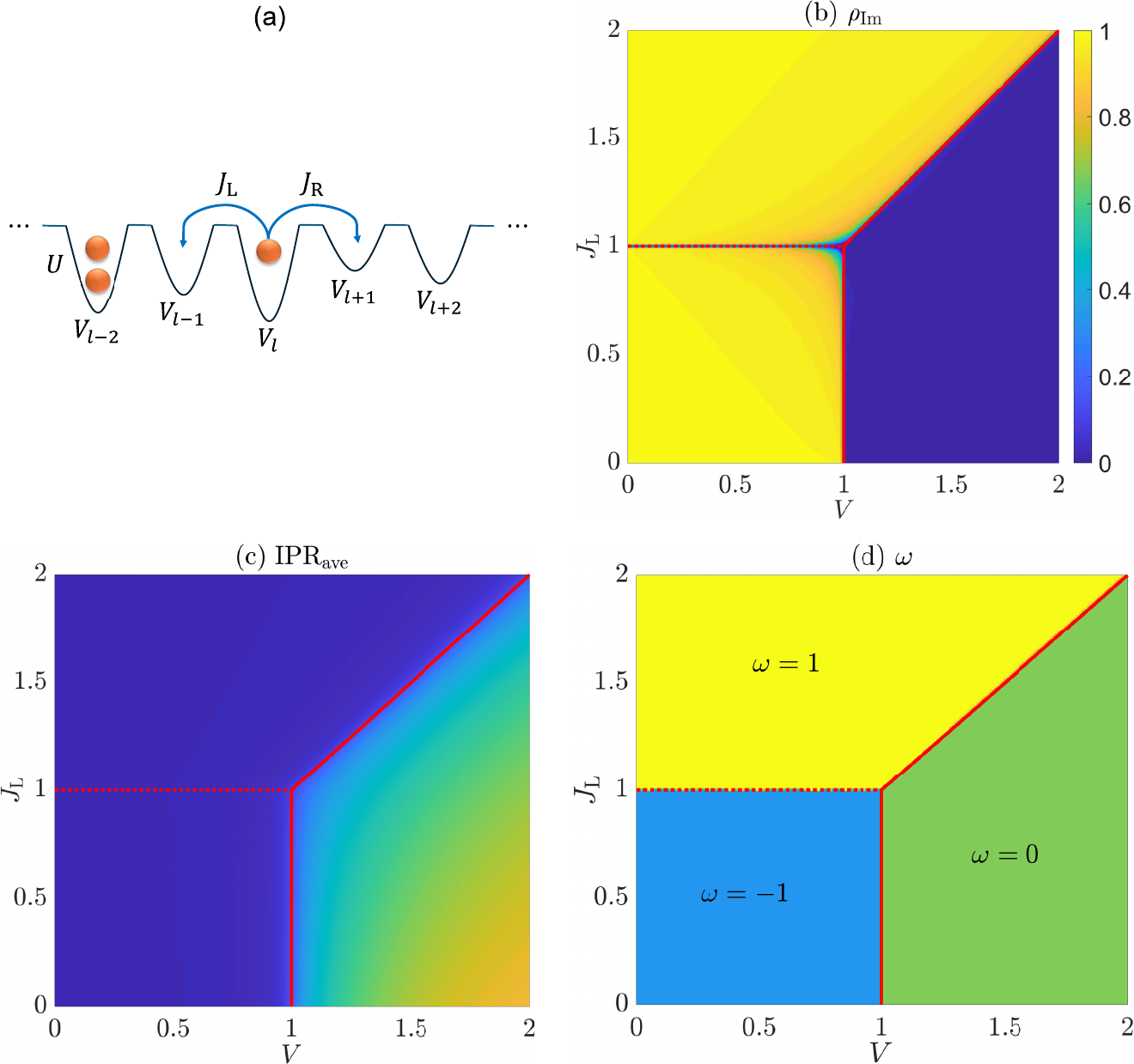}
\par\end{centering}
\caption{A sketch of the model {[}in (a){]} and its single-particle
phases described by different variables {[}in (b)--(d), with
$U=0${]} under the PBC. In (a), we have $V_{l}\equiv2V\cos(2\pi\alpha l)$,
which is quasiperiodic in $l$ for any irrational $\alpha$. (b) shows
the ratio of eigenenergies of $\hat{H}_{0}$ with nonzero imaginary
parts $\rho_{{\rm Im}}$ [Eq.~(\ref{eq:RIm})] vs $(V,J_{{\rm L}})$. (c) shows the averaged
IPRs [Eq.~(\ref{eq:IPRa})] over all eigenstates of $\hat{H}_{0}$ vs $(V,J_{{\rm L}})$.
(b) and (c) share the same color bar. (d) gives the topological phase
diagram characterized by the winding number $\omega$ [Eq.~(\ref{eq:w})]. In (b)--(d),
the solid line denotes the boundary {[}Eq.~(\ref{eq:PB}){]} of PT,
localization and topological transitions under the PBC, while the
dotted line represents the boundary ($J_{{\rm L}}=J_{{\rm R}}$ and
$J_{{\rm L}},J_{{\rm R}}<V$) between left and right skin-localizations
of bulk states under the OBC. The hopping amplitude $J_{{\rm R}}=1$
and the length of lattice $L=89$ are set to be the same for (b)--(d).
\label{fig:Sketch}}
\end{figure}

In this work, we investigate phases and transitions caused by the
collaboration among quasiperiodic disorder, non-Hermitian asymmetric
couplings and interparticle interactions in one dimension. A minimal model that takes into account all these competing factors
can be described by the tight-binding Hamiltonian
\begin{alignat}{1}
\hat{H}= & \sum_{l}\left(J_{{\rm L}}\hat{b}_{l}^{\dagger}\hat{b}_{l+1}+J_{{\rm R}}\hat{b}_{l+1}^{\dagger}\hat{b}_{l}\right)\nonumber \\
+ & \sum_{l}\left[2V\cos(2\pi\alpha l)\hat{n}_{l}+\frac{U}{2}\hat{n}_{l}(\hat{n}_{l}-1)\right].\label{eq:H}
\end{alignat}
A schematic diagram of the model is shown in Fig.~\ref{fig:Sketch}(a).
$\hat{b}_{l}^{\dagger}$ ($\hat{b}_{l}$) creates (annihilates) a
boson on the lattice site $l\in\mathbb{Z}$. $\hat{n}_{l}=\hat{b}_{l}^{\dagger}\hat{b}_{l}$
is the particle number operator on the $l$th site. $J_{{\rm L}}$
and $J_{{\rm R}}$ are right-to-left and left-to-right nearest-neighbor
hopping amplitudes, respectively, which are asymmetric so long as
$J_{{\rm L}}^{*}\neq J_{{\rm R}}$. Without loss of generality, we
will only consider cases with $J_{{\rm L}},J_{{\rm R}}\geq0$. The Hamiltonian $\hat{H}$ is then non-Hermitian ($\hat{H}\neq\hat{H}^{\dagger}$)
if $J_{{\rm L}}\neq J_{{\rm R}}$. $V\in\mathbb{R}$ controls the
strength of onsite potential. The latter is quasiperiodic when $\alpha$
is irrational. Throughout the calculations, we choose $\alpha$ to
be the inverse golden ratio $\frac{\sqrt{5}-1}{2}$.
Taking other irrational values for $\alpha$ yields consistent
results. Finally, $U$ is the strength of onsite Bose-Hubbard
interactions, i.e., the interaction energy of two bosons on the same
lattice site.

Without interactions, the system described by Eq.~(\ref{eq:H}) reduces
to the quasiperiodic nonreciprocal AAH model, whose properties has
been investigated \cite{jiang2019interplay,longhi2021phase}. Let us denote the noninteracting
part of $\hat{H}$ by $\hat{H}_{0}\equiv\hat{H}(U=0)$. Under the
periodic boundary condition (PBC), a three-fold transition was identified
at
\begin{equation}
|V|=\max(J_{{\rm L}},J_{{\rm R}}).\label{eq:PB}
\end{equation}
When $|V|<\max(J_{{\rm L}},J_{{\rm R}})$, the spectrum of $\hat{H}_{0}$
is complex and all its eigenstates are spatially extended. When $|V|>\max(J_{{\rm L}},J_{{\rm R}})$,
the spectrum of $\hat{H}_{0}$ becomes real and all its eigenstates
are exponentially localized. Moreover, a spectral winding number $\omega$
can be defined for $\hat{H}_{0}$, whose value is $\omega=\pm1$ ($\omega=0$)
when $\hat{H}_{0}$ holds a complex (real) spectrum. Therefore, at
the triple criticality {[}Eq.~(\ref{eq:PB}){]} of the noninteracting
$\hat{H}_{0}$, we encounter a PT transition in the spectrum,
a localization transition of the eigenstates, plus a topological
transition accompanied by the unit jump of winding number $\omega$.
Meanwhile, the system described by $\hat{H}_{0}$ has NHSEs under
OBC due to the hopping asymmetry when
$J_{{\rm L}}\neq J_{{\rm R}}$. Interestingly, the winding number
$\omega$ takes the value $-1$ ($+1$) when $J_{{\rm L}}<J_{{\rm R}}$
($J_{{\rm L}}>J_{{\rm R}}$). The sign of $\omega$ may then be used
to distinguish the direction of skin-localization, which is towards
the left (right) edge if $J_{{\rm L}}<J_{{\rm R}}$ ($J_{{\rm L}}>J_{{\rm R}}$).
In comparison, the triple criticality is reduced to a single localization transition
point and the NHSEs disappear in the Hermitian limit ($J_{{\rm L}}=J_{{\rm R}}$) \cite{aubry1980analyticity}.
The physics of quasiperiodic AAH model could thus be greatly enriched
by nonreciprocal non-Hermitian effects.

In Figs.~\ref{fig:Sketch}(b)--\ref{fig:Sketch}(d), we illustrate
the phases and transitions of the noninteracting system via computing
some relevant quantities, including the ratio of complex eigenenergies
$\rho_{{\rm Im}}$, the mean ${\rm IPR}_{{\rm ave}}$ of IPRs 
and the winding number $\omega$. These quantities are defined below. Let us denote the total number of energy eigenstates by $D$
and the number of states whose eigenenergies have nonzero imaginary
parts by $D_{{\rm Im}}$. The $\rho_{{\rm Im}}$ is then given by 
\begin{equation}
\rho_{{\rm Im}}=D_{{\rm Im}}/D.\label{eq:RIm}
\end{equation}
When $\rho_{{\rm Im}}=0$ ($\rho_{{\rm Im}}>0$), the system resides
in the PT invariant (PT broken) region with a real (complex) spectrum.
These two regions are separated by the solid line in 
Fig.~\ref{fig:Sketch}(b), which satisfies the phase boundary condition
Eq.~(\ref{eq:PB}) of the noninteracting $\hat{H}_{0}$. Let
$|\psi_{j}\rangle$ be the $j$th right eigenstate of the system's
Hamiltonian with energy $E_{j}$ ($j=1,...,D$). We denote the overlap
between $|\psi_{j}\rangle$ and the $n$th Fock basis by $\psi_{n}^{(j)}$
($n=1,...,D$). The IPR of $|\psi_{j}\rangle$ is then given by 
\begin{equation}
{\rm IPR}_{j}=\sum_{n=1}^{D}|\psi_{n}^{(j)}|^{4}/(\sum_{n=1}^D|\psi^{(j)}_n|^2)^2.\label{eq:IPRj}
\end{equation}
Averaging ${\rm IPR}_{j}$ over all the eigenstates $|\psi_{j}\rangle$ yields
\begin{equation}
{\rm IPR}_{{\rm ave}}=\frac{1}{D}\sum_{j=1}^{D}{\rm IPR}_{j}.\label{eq:IPRa}
\end{equation}
Note in passing that we only use right eigenvectors and their standard
normalization convention in the calculation of IPR. The IPR of each eigenstate then takes values in the range $[0,1]$.
In the limit of large system size $L\rightarrow\infty$, we have ${\rm IPR}_{{\rm ave}}\rightarrow0$
if all the eigenstates $|\psi_{j}\rangle$ are spatially extended, since the IPR of each state goes to zero as $1/L$ in this case.
On the other hand, if a significant potion $(\propto D)$ of the eigenstates
are localized, the ${\rm IPR}_{{\rm ave}}$ would take a finite value, as the IPR of each localized state is finite in the limit $L\rightarrow\infty$.
In Fig.~\ref{fig:Sketch}(c), we find a clear boundary between the
regions with vanishing and finite ${\rm IPR}_{{\rm ave}}$, which
is given by Eq.~(\ref{eq:PB}) for the noninteracting $\hat{H}_{0}$.
The same boundary can be identified by evaluating the maximum and
minimum of IPRs, which confirms that Eq.~(\ref{eq:PB}) indeed describes
the critical line of localization transition in the quasiperiodic
nonreciprocal AAH model. Finally, the winding number $\omega$ counts
the number of times that the spectrum of the system winds around
a base point $E_{0}^{{\rm B}}$ on the complex energy plane. For the noninteracting
$\hat{H}_{0}$, it can be expressed as
\begin{equation}
\omega=\int_{0}^{2\pi}\frac{d\phi}{2\pi i}\partial_{\phi}\ln\{\det[\hat{H}_{0}(\phi/L)-E_{0}^{{\rm B}}]\}.\label{eq:w}
\end{equation}
Here, $L$ is the length of lattice and we choose $E_{0}^{{\rm B}}=0$
in numerical calculations. The phase factor $\phi/L$ is introduced
via adding a synthetic flux $\phi$ over the system under the PBC
and distributing it uniformly across each bond, so that the hopping
amplitudes of the chain are modified as $J_{{\rm L}}\rightarrow J_{{\rm L}}e^{i\phi/L}$
and $J_{{\rm R}}\rightarrow J_{{\rm R}}e^{-i\phi/L}$. It is clear
that we have $\omega=0$ if the spectrum of the system is real. If
the spectrum is complex with a single loop around $E_{0}^{{\rm B}}$,
we would have $\omega=\pm1$. These cases are observed in Fig.~\ref{fig:Sketch}(d). Across the PT and localization
transitions {[}Eq.~(\ref{eq:PB}){]}, $\omega$ goes from
$\pm1$ to $0$. These transitions are thus topological. Passing through
the dotted line in Fig.~\ref{fig:Sketch}(d), $\omega$ goes
from $-1$ to $1$, which signifies the change of skin-localization
direction between left and right edges of the chain under the OBC.
Note in passing that no changes are observed around the same dotted
line in Fig.~\ref{fig:Sketch}(c), which is due to the disappearance
of NHSEs and skin-localizations under the PBC.

To sum up, nonreciprocal non-Hermitian effects could
extend the localization transition point of quasiperiodic AAH model
{[}at $J_{\rm L}=V=1$ in Figs.~\ref{fig:Sketch}(b)--\ref{fig:Sketch}(d){]}
into critical lines, through which the system could undergo PT, localization
and topological triple phase transitions under the PBC or reverse
its skin-localization direction under the OBC. With interactions,
the critical point could further expand into a whole
intermediate phase, showing both localized and extended states separated by mobility edges under the
PBC and skin-localized doublons under the OBC, as will be shown
in the following section. Before that, we introduce our tools
to characterize the phases and transitions of two interacting
bosons.
	
\subsection{Methods}\label{sec.method}
With interactions, we could also identify the spectral transition
of $\hat{H}$ by the ratio of states whose imaginary part of eigenenergies
are nonzero, i.e., the $\rho_{{\rm Im}}$ in Eq.~(\ref{eq:RIm}),
where $D=\frac{(N+L-1)!}{N!(L-1)!}$ counts the dimension of Fock
space with two bosons ($N=2$) in a lattice of $L$ sites. We would
have a real spectrum when $\rho_{{\rm Im}}=0$, a mixed one
with both real and complex eigenenergies when $0<\rho_{{\rm Im}}<1$,
and a fully complex spectrum when $\rho_{{\rm Im}}\simeq1$.

In our exact diagonalization calculations, we express the Hamiltonian in Fock
space with the occupation number representation $\{|2_l\rangle,l=1,...,L\}$ and $\{|1_{l},1_{m}\rangle|l,m=1,...,L;\,l<m\}$.
Here, $|2_l\rangle$ ($|1_{l},1_{m}\rangle$) denotes the Fock basis with two bosons on the same (different) lattice site $l$ (sites $l$ and $m$).
To characterize the localization nature of states in real space, we
define the symmetrized position eigenbasis as the union of sets $\{|l,l\rangle\}$
and $\{(|l,m\rangle+|m,l\rangle)/\sqrt{2}\}$, where $l,m=1,2,...,L$
and $l\neq m$.
Here, $|l,m\rangle\equiv|l\rangle_1\otimes|m\rangle_2$ is just the direct product state of
position eigenbases in the Hilbert spaces of the two bosons.
The first and second indices of each ket $|l,m\rangle$ denote the
positions of two bosons in the lattice.
The transformations between Fock and position bases are given by
$|2_l\rangle=|l,l\rangle$ and $|1_l,1_m\rangle=(|l,m\rangle+|m,l\rangle)/\sqrt{2}$.
The IPR and its average in
Eqs.~(\ref{eq:IPRj}) and (\ref{eq:IPRa}) can then be computed with
respect to the symmetrized position eigenbasis after exact diagonalization. To figure out the existence of intermediate
phases with both extended and localized states, we further introduce
the maximum and minimum of IPRs as
\begin{alignat}{1}
{\rm IPR}_{\max} & =\max_{j\in\{1,...,D\}}({\rm IPR}_{j}),\label{eq:IPRmax}\\
{\rm IPR}_{\min} & =\min_{j\in\{1,...,D\}}({\rm IPR}_{j}).\label{eq:IPRmin}
\end{alignat}
In the limit of large system size $L\rightarrow\infty$, we have both
${\rm IPR}_{\max}\rightarrow0$ and ${\rm IPR}_{\min}\rightarrow0$
in the extended phase, ${\rm IPR}_{\max}>0$ and ${\rm IPR}_{\min}>0$
in the localized phase, and only ${\rm IPR}_{\min}\rightarrow0$ in
the intermediate mobility edge phase. Collecting the information provided by ${\rm IPR}_{\max}$
and ${\rm IPR}_{\min}$ then allows us to distinguish the extended,
intermediate and localized phases. We may also use a single quantity
$\zeta$ to identify the presence of intermediate mobility edge phases, which is given by \cite{Sarma2020PRB}
\begin{equation}
\zeta=\log_{10}({\rm IPR}_{{\rm ave}}\cdot{\rm NPR}_{{\rm ave}}),\label{eq:zeta}
\end{equation}
where ${\rm NPR}_{{\rm ave}}\equiv\sum_{j=1}^{D}{\rm IPR}_{j}^{-1}/(LD)$
is the average of normalized participation ratios. In the limit
$L\rightarrow\infty$, the function $\zeta$ takes finite
values in the intermediate mobility edge phase while approaching $-\infty$ in both
extended and localized phases. We could then employ ${\rm IPR}_{\max}$,
${\rm IPR}_{\min}$ and $\zeta$ to fully characterize the localization
nature of different phases and localization transitions in our system.
Note in passing that for better illustrations, our presented values of $\zeta$ are shifted and rescaled as $(\zeta-\min(\zeta))/\max(\zeta-\min(\zeta))$, where the $\min(x)$ and $\max(x)$ find the minimum and maximum of $x$ over the considered domain of system parameters in Figs.~\ref{fig:UHMfig2} and \ref{fig:AHMfig2} of the next section.

\begin{figure}
\begin{centering}
\includegraphics[scale=0.48]{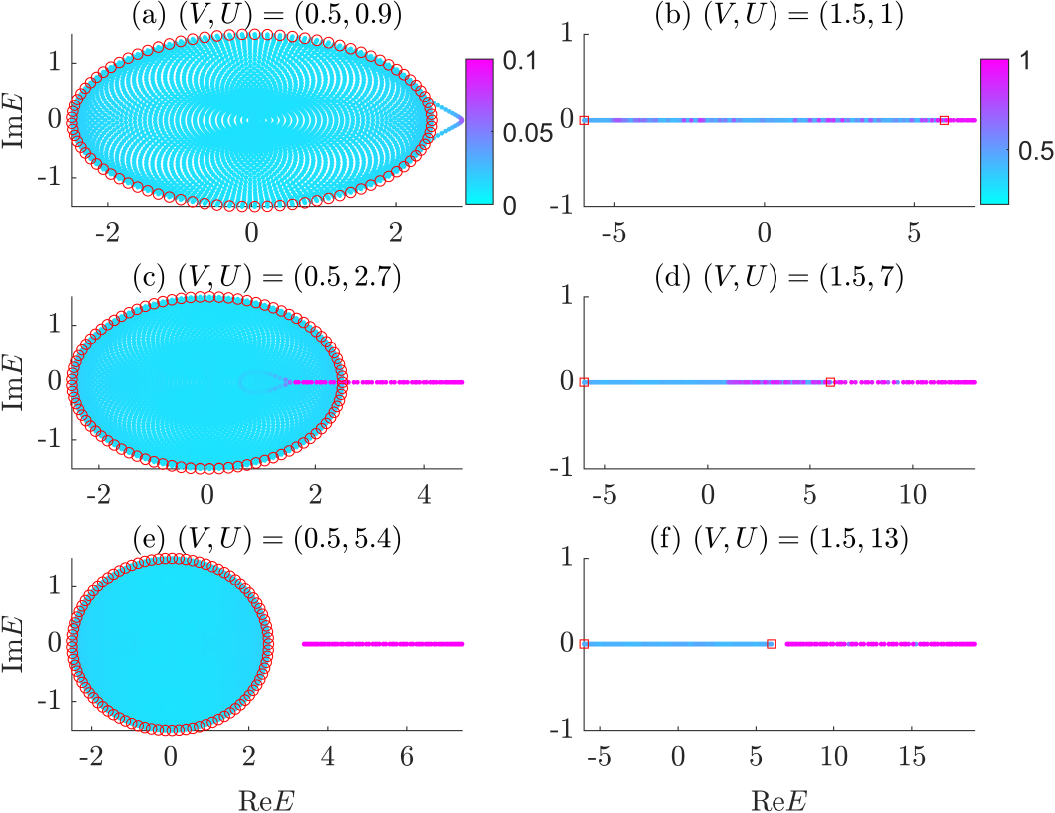}
\par\end{centering}
\caption{Energy spectra of the UHM under the PBC. The hopping amplitude is
$J=1$ and the length of lattice is $L=89$ for all panels. The IPR
of each state is given by the color bar. The circles in (a), (c) and
(e) denote the spectral boundary $E_{{\rm b}}$ {[}Eq.~(\ref{eq:Eb}){]}
of the noninteracting UHM with $V<J$. The squares in (b), (d), and
(f) denote the spectral ends $E=\pm4V$ of the noninteracting UHM
with $V>J$. (a), (c) and (e) [(b), (d) and (f)] share the same color bar. \label{fig:UHMfig1}}
\end{figure}

To identify the topological signatures of spectral and localization
transitions with two interacting bosons, we note that the single winding
number in Eq.~(\ref{eq:w}) may not be enough, as the extended-to-intermediate
and intermediate-to-localized transitions may possess different phase
boundaries. To resolve this issue, we introduce a pair of winding
numbers $(\omega_{1},\omega_{2})$, defined versus two different base
energies $(E_{1}^{{\rm B}},E_{2}^{{\rm B}})$ as
\begin{equation}
\omega_{\ell}=\int_{0}^{2\pi}\frac{d\phi}{2\pi i}\partial_{\phi}\ln\{\det[\hat{H}(\phi/L)-E_{\ell}^{{\rm B}}]\},\quad\ell=1,2,\label{eq:w12}
\end{equation}
where the parametrization of $\hat{H}(\phi/L)$ from $\hat{H}$ is
the same as that of $\hat{H}_{0}(\phi/L)$ {[}see the discussion below
Eq.~(\ref{eq:w}){]}. In numerical calculations, we choose $E_{1}^{{\rm B}}=0$
and set $E_{2}^{{\rm B}}$ as the real part of energy of the eigenstate
of $\hat{H}$ with the largest IPR. The quantized change of $\omega_{1}$
from zero to a finite integer then signifies the transition of
the spectrum from real to complex. Meanwhile, the quantized jump of
$\omega_{2}$ from zero to a finite integer corresponds to the delocalization
transition, after which all the eigenstates become extended. Therefore,
we would have $\omega_{1}=\omega_{2}=0$ in a phase with real spectrum
and only localized states, $\omega_{1},\omega_{2}\neq0$ in a phase
with complex spectrum and only extended states, and $\omega_{1}\neq0$,
$\omega_{2}=0$ in an intermediate mobility edge phase with both extended and localized eigenstates. Topological signals of the phase transitions could then be
captured by the changes of $\omega_{1}$ or $\omega_{2}$ between
zero and finite values.

In the next section, we reveal and characterize the interaction induced
spectral transitions, localization transitions, and intermediate phases in our
system from an integrated perspective of eigenspectrum, state distributions and topology, with the help of the tools introduced in this section.

\section{Interaction-driven phases and transitions}\label{sec.result}
In this section, we demonstrate the effect of interactions on NHQCs by investigating two representative nonreciprocal hopping
models. Their spectral, localization, and topological properties will
be treated separately in Secs.~\ref{sec.unidirect} and \ref{sec.bidirect}, with a focus on two interacting bosons. Doublons emerging in strong interaction regions and their NHSEs will be discussed in the next section.

\subsection{Unidirectional hopping Model (UHM)}\label{sec.unidirect}
We begin our study of the two-body problem with a minimal non-Hermitian AAH model, in
which hopping terms along only one spatial direction are kept. Without
losing generality, we set $J_{{\rm L}}=0$ and $J_{{\rm R}}=J$ in
Eq.~(\ref{eq:H}), so that particles can only hop from left to right.
The Hamiltonian of this UHM reads
\begin{equation}
\hat{H}_{1}=\sum_{l}\left[J\hat{b}_{l+1}^{\dagger}\hat{b}_{l}+2V\cos(2\pi\alpha l)\hat{n}_{l}+\frac{U}{2}\hat{n}_{l}(\hat{n}_{l}-1)\right].\label{eq:H1}
\end{equation}

In cases with $U=0$, the spectrum $E$ of $\hat{H}_{1}$ with two
bosons can be deduced by adding their single-particle eigenenergies
\cite{longhi2021phase}. Under the PBC, we would obtain
\begin{equation}
E=\begin{cases}
2V(\cos k+\cos k'), & J\leq V,\\
J(e^{-ik}+e^{-ik'})+\frac{V^{2}}{J}(e^{ik}+e^{ik'}), & J>V.
\end{cases}\label{eq:H1E}
\end{equation}
Here, $k$ and $k'$ are two synthetic parameters, each filling the domain $[-\pi,\pi)$ densely in the large-size limit $L\rightarrow\infty$.
When $J<V$, the spectrum is real with two ending points at $E=\pm4V$,
and all the eigenstates of $\hat{H}_{1}$ are localized. When $J>V$,
the spectrum is complex with eigenenergies filling an ellipse.
Its outer boundary, according to the noninteracting spectrum Eq.~(\ref{eq:H1E}), is located at
\begin{equation}
E_{{\rm b}}=2Je^{-ik}+2(V^{2}/J)e^{ik},\quad k\in[-\pi,\pi).\label{eq:Eb}
\end{equation}
All the eigenstates of $\hat{H}_{1}$ are extended in this case. Thus, a real-complex spectral transition and a localization transition
occur together at $J=V$. In the following, we consider the
system with two interacting bosons and set $J=1$ as the unit of energy.

When $U\neq0$, the spectra in Eq.~(\ref{eq:H1E}) are modified by
the interaction between two bosons. The eigensystem of $\hat{H}_{1}$
can then be obtained numerically by exact diagonalization. In 
Fig.~\ref{fig:UHMfig1}, we present the spectra of $\hat{H}_{1}$ and the
IPRs of the associated eigenstates for some typical cases. We find
that for $V<J$, a majority of the eigenenergies are complex and confined
within the boundary described by Eq.~(\ref{eq:Eb}). Meanwhile, with
the increase of $U$, some eigenvalues are first deformed away from
the main part of spectrum, forming a small loop around the highest
energy of the noninteracting system, and then collapsed onto the ${\rm Re}E$
axis. Interestingly, the states with real energies generated by a
nonzero $U$ are all spatially localized. We thus obtain an interaction-driven
intermediate mobility edge phase, in which extended and localized eigenstates coexist,
which is absent in the noninteracting UHM. The further increase of
$U$ will generate at most $L$ localized eigenstates with real energies.
When $U$ is large enough, their spectrum would finally be gapped
from the extended states with complex energies inside $E_{{\rm b}}$.
Each eigenstate in the large $U$ regime is formed by a bosonic doublon
localized around one lattice site. When $V>J$, the spectrum of $\hat{H}_{1}$
retains to be real and all its eigenstates are localized for any nonzero
$U$. This is not hard to expect, as the onsite Bose-Hubbard interaction
could only enhance localization, and the localized particles mainly
feel the Hermitian quasiperiodic onsite potential $2V\cos(2\pi\alpha l)$.
At large $U$, the spectrum splits into two parts, which are separated
by a gap along the ${\rm Re}E$ axis. The collection of states with
higher energies are again doublons, with each of them containing two
bosons localized on the same lattice site.

\begin{figure}
\begin{centering}
\includegraphics[scale=0.48]{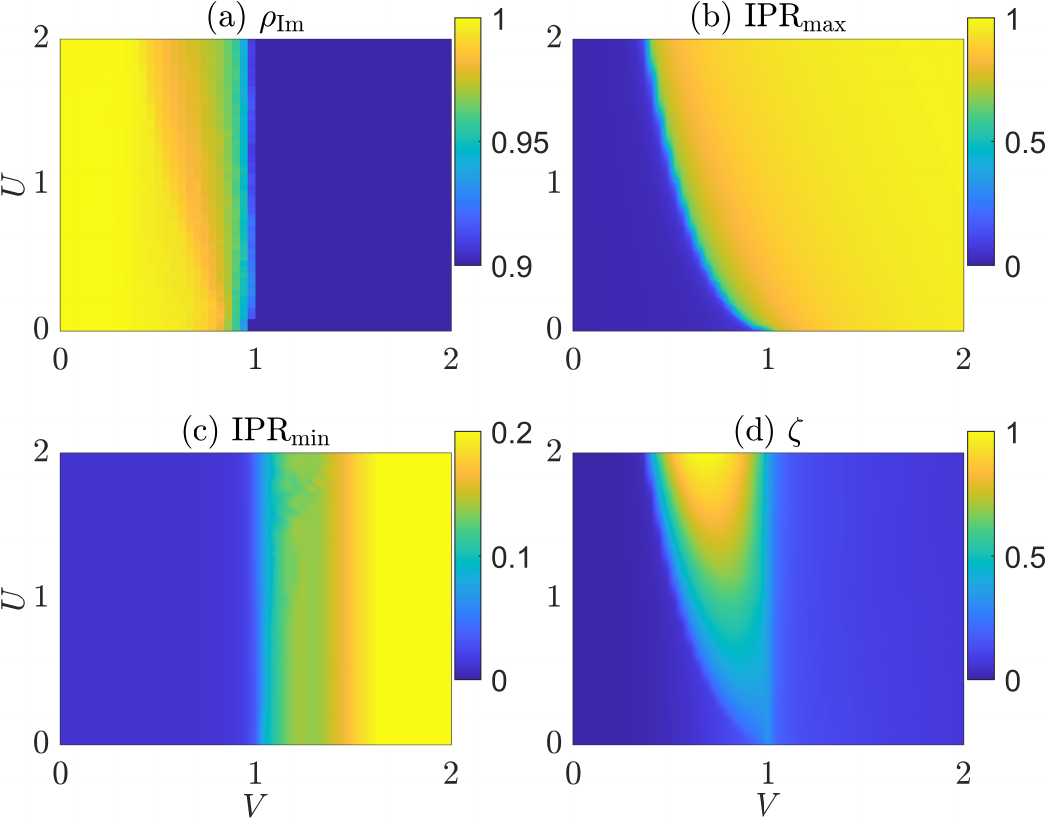}
\par\end{centering}
\caption{Spectral and localization transitions of the UHM vs $(V,U)$ under
the PBC. The hopping amplitude is $J=1$ and the length of lattice
is $L=89$ for all panels. (a) Ratio of eigenenergies with nonzero
imaginary parts $\rho_{{\rm Im}}$ {[}Eq.~(\ref{eq:RIm}){]}.
(b) The maximum of IPRs ${\rm IPR}_{\max}$ {[}Eq.~(\ref{eq:IPRmax}){]}.
(c) The minimum of IPRs ${\rm IPR}_{\min}$ {[}Eq.~(\ref{eq:IPRmin}){]}.
(d) The smoking-gun function $\zeta$ {[}Eq.~(\ref{eq:zeta}){]} of
the intermediate mobility edge phase. \label{fig:UHMfig2}}
\end{figure}

The above results imply that interactions could transform part of
the extended phase of the free-boson UHM into a intermediate one, which
contains both extended two-boson states and localized doublons when
the interaction is strong enough. To further confirm this, we
present the characteristic functions $\rho_{{\rm Im}}$ {[}Eq.~(\ref{eq:RIm}){]},
${\rm IPR}_{\max}$ {[}Eq.~(\ref{eq:IPRmax}){]}, ${\rm IPR}_{\min}$
{[}Eq.~(\ref{eq:IPRmin}){]} and $\zeta$ {[}Eq.~(\ref{eq:zeta}){]}
of spectra and states over a range of onsite potential and interaction
strengths, as shown in Fig.~\ref{fig:UHMfig2}. In 
Fig.~\ref{fig:UHMfig2}(a), we observe a complex-to-real spectral transition
at $V=J=1$. The same boundary is found in Fig.~\ref{fig:UHMfig2}(c),
across which all eigenstates become localized. These observations
suggest that the final accomplishment of complex-to-real spectral
transition and localization transition are not affected by the onsite
Bose-Hubbard interaction in our UHM. However, we notice in  Fig.~\ref{fig:UHMfig2}(b)
a boundary between the regions with ${\rm IPR}_{\max}\simeq0$ and
${\rm IPR}_{\max}>0$, which is different from the boundary $V=J$
of ${\rm IPR}_{\min}$ in Fig.~\ref{fig:UHMfig2}(c) while coinciding
with the boundary between $\rho_{{\rm Im}}\simeq1$ and $0<\rho_{{\rm Im}}<1$
in Fig.~\ref{fig:UHMfig2}(a). This new phase boundary corresponds
to the onset of spectral and localization transitions, which emerges
only when $U\neq0$ and thus has an interaction-origin. However, it
could not transform all the states from extended to localized, and
the whole spectrum from complex to real. In the region between $V=J$
and the boundary in Fig.~\ref{fig:UHMfig2}(b), we now have an intermediate
phase with coexisting localized states of real energies and extended
states of complex energies. Such a phase is absent without interactions,
and its appearance is further confirmed by the function $\zeta$ in
Fig.~\ref{fig:UHMfig2}(d), which shows positive values only in the
intermediate mobility edge phase. Putting together, we confirm that interactions in
the UHM $\hat{H}_{1}$ could indeed generate a mobility edge phase and
drive the boundary of extended phase towards the region with weaker
disorder strengths.

\begin{figure}
\centering
\includegraphics[width=0.95\linewidth]{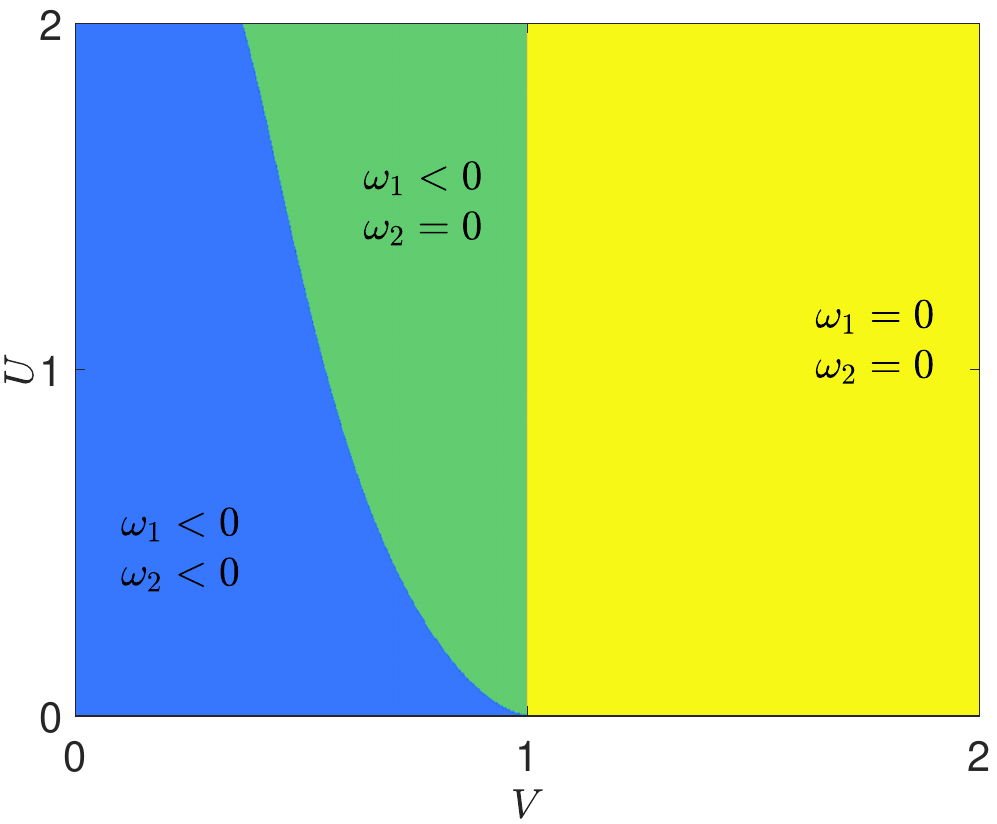}
\caption{Phase diagram of the UHM vs $(V,U)$ under
the PBC, characterized by the winding numbers $(\omega_1,\omega_2)$ [Eq.~(\ref{eq:w12})]. The hopping amplitude is $J=1$ and the length of lattice is $L=89$.}
\label{fig:UHMfig3}
\end{figure}

To decode the topological nature of the observed spectral and localization
transitions, we compute the winding numbers $(\omega_{1},\omega_{2})$ introduced
in Eq.~(\ref{eq:w12}), yielding the phase diagram of the UHM in 
Fig.~\ref{fig:UHMfig3}. Within the localized phase, the spectrum of $\hat{H}_{1}$ is real
and we find $\omega_{1}=\omega_{2}=0$, as expected. In the intermediate mobility edge phase,
we obtain $\omega_{1}<0$ and $\omega_{2}=0$. The value of $\omega_{1}$ then undergoes
a quantized jump through the real-to-complex spectral transition at
$V=J$. In the extended phase, we find both $\omega_{1}<0$ and $\omega_{2}<0$.
The value of $\omega_{2}$ thus experiences a quantized change at the boundary
through which all the eigenstates become extended. Therefore, the
locations where $\omega_{1}$ and $\omega_{2}$ start to deviate from zero in
the parameter space could correctly capture all the phase boundaries,
offering topological signatures of phase transitions in our system.
Note in passing that the exact integer values of $\omega_{1}$ and $\omega_{2}$
rely on the times that the corresponding base energies $E^{\rm B}_{1}$ and
$E^{\rm B}_{2}$ are encircled {[}see Eq.~(\ref{eq:w12}){]}, which could
be further dependent on the system size when $\omega_{1}\neq0$ and/or
$\omega_{2}\neq0$. For the lattice size $L=89$ used in our calculations,
we find $\omega_{2}=-2$ ($=0$) in the extended (intermediate) phase and $\omega_{1}=-27$
in the extended/intermediate phase, respectively. Therefore, the winding numbers
$(\omega_{1},\omega_{2})$ are expected to provide topological signatures of
phase transitions in our system, rather than topological characterizations
of the phases with different spectral and localization properties.

To sum up, we find that in the UHM, the interaction between two bosons could enlarge the critical
point of the noninteracting system at $V=J$ to an intermediate mobility edge phase,
possessing both extended and localized two-boson states with complex
and real energies, respectively. The transition from localized
to intermediate phases also accompanies a real-to-complex spectral transition,
while the transition to extended phase is moved up to the domain with
weaker disorder in the presence of interactions. In Sec.~\ref{sec.bidirect},
we consider the impact of interactions on the more general AAH NHQC
with asymmetric hopping amplitudes.
  
\subsection{Asymmetric hopping model (AHM)}\label{sec.bidirect}

\begin{figure}
\begin{centering}
\includegraphics[scale=0.48]{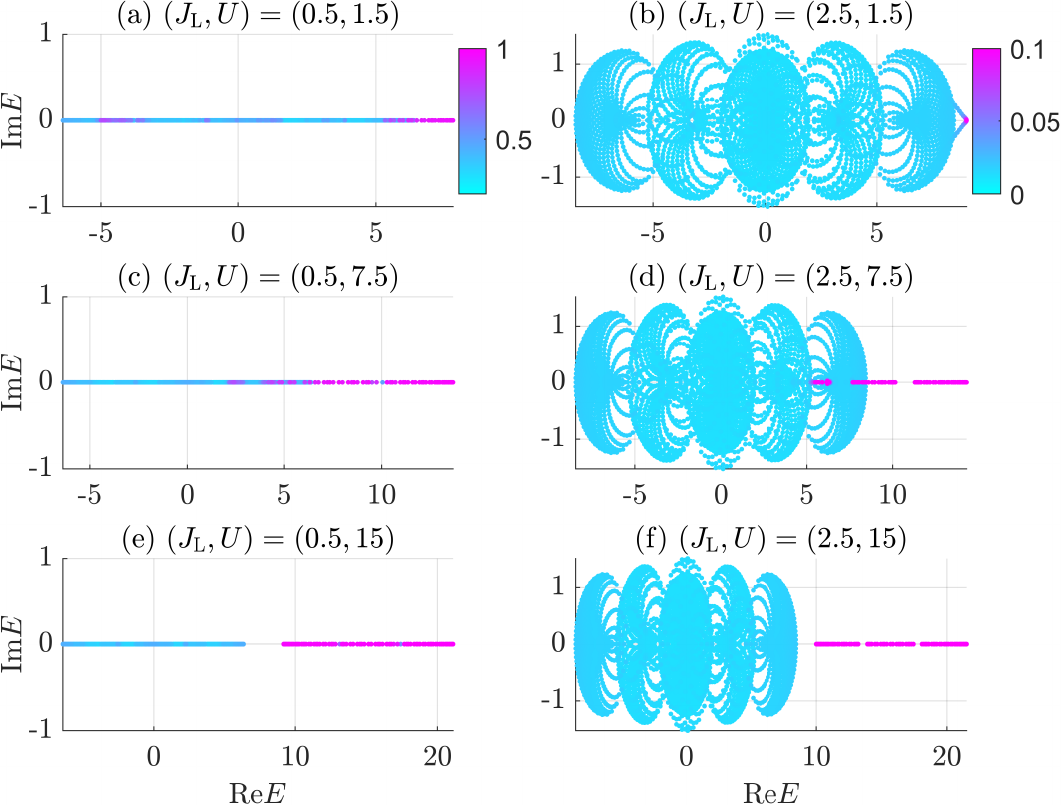}
\par\end{centering}
\caption{Energy spectra of the AHM under the PBC. Other system parameters are
$(J_{{\rm R}},V)=(1,1.5)$ and the length of lattice is $L=89$ for
all panels. The IPR of each state is given by the color bar. (a),
(c) and (e) {[}(b), (d) and (f){]} share the same color bar. \label{fig:AHMfig1}}
\end{figure}

We now go back the original AHM in Eq.~(\ref{eq:H}), treating the
case with $J_{{\rm L}},J_{{\rm R}}\neq0$ and $J_{{\rm L}}\neq J_{{\rm R}}$.
To have a good comparison with the results of noninteracting
model (as discussed in Sec.~\ref{sec.GM}), we set $J_{{\rm R}}=1$ as the
unit of energy and let the quasiperiodic potential strength $V=1.5$ throughout this subsection.
In this case, we have a real-to-complex spectral transition and also
a localization transition at $J_{{\rm L}}=V=1.5$ when
$U=0$, as shown in Figs.~\ref{fig:Sketch}(b)--\ref{fig:Sketch}(d).
Meanwhile, even without interactions, simple expressions of the spectra
under PBC could not be found in this case \cite{longhi2021phase}.
We thus resort to numerical calculations for characterizing the spectra
and states of the AHM.

In Fig.~\ref{fig:AHMfig1}, we present the spectra of $\hat{H}$ [Eq.~(\ref{eq:H})] and
the IPRs of its eigenstates for typical cases with $U\neq0$. Qualitatively,
we see that the spectral and state features of the UHM are maintained
even with nonzero $J_{{\rm L}}$ and $J_{{\rm R}}$. Specially, we
find that when $J_{{\rm L}}<V$, the spectrum of $\hat{H}$ is real
and all its eigenstates are localized for $U\neq0$. Since the onsite
interaction only promotes localization, and localized bosons are controlled
by the quasiperiodic potential, the spectral
properties observed in Figs.~\ref{fig:AHMfig1}(a), \ref{fig:AHMfig1}(c)
and \ref{fig:AHMfig1}(e) are expected. At large $U$, the spectrum
also divides into two portions gapped along ${\rm Re}E$ as that of
UHM. The states with higher energies are bosonic doublons localized
at different lattice sites. For $J_{{\rm L}}>V$, large amounts of
eigenenergies become complex. With the rise of $U$, some energies start
to leave the main body of spectrum, composing a loop at the highest
energy of the noninteracting AHM, and then pinned to ${\rm Re}E$.
Similar to UHM, the states with real energies due to a nonzero $U$
are spatially localized. We should then have an interaction-induced intermediate mobility edge
phase in the AHM with coexisting extended and localized eigenstates,
which disappears when $U=0$. Further increasing $U$ yields $L$
localized states with real energies. Finally, the spectrum becomes
two disconnected sets containing extended states with complex energies
and localized bosonic doublons with real energies, respectively. These
observations suggest that generic features of the UHM regarding spectral,
localization and topological transitions should be carried over to
the AHM with both $J_{{\rm L}},J_{{\rm R}}\neq0$.

\begin{figure}
\begin{centering}
\includegraphics[scale=0.475]{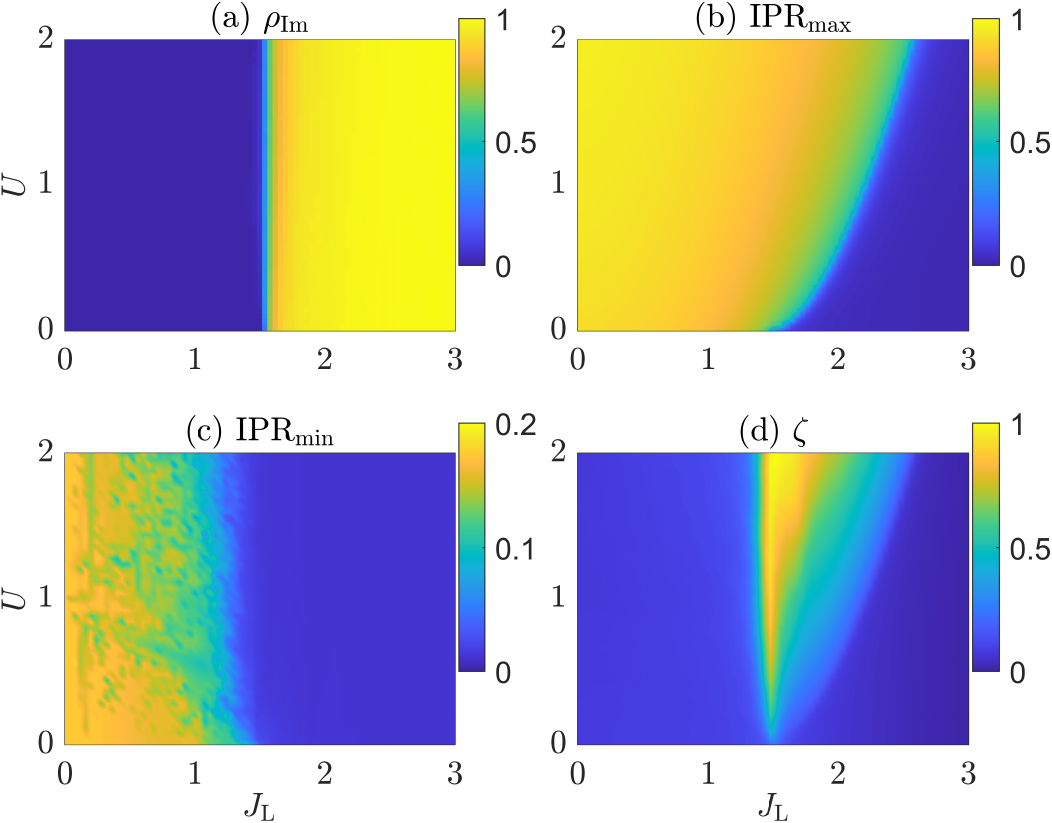}
\par\end{centering}
\caption{Spectral and localization transitions of the AHM vs $(J_{{\rm L}},U)$
under the PBC. Other system parameters are $(J_{{\rm R}},V)=(1,1.5)$
and the length of lattice is $L=89$ for all panels. (a) Ratio of
eigenenergies with nonzero imaginary parts $\rho_{{\rm Im}}$ {[}Eq.~(\ref{eq:RIm}){]}. (b) The maximum of IPRs ${\rm IPR}_{\max}$ {[}Eq.~(\ref{eq:IPRmax}){]}. (c) The minimum of IPRs ${\rm IPR}_{\min}$
{[}Eq.~(\ref{eq:IPRmin}){]}. (d) The smoking-gun function $\zeta$
{[}Eq.~(\ref{eq:zeta}){]} of the intermediate mobility edge phase.
\label{fig:AHMfig2}}
\end{figure}

To gain a more comprehensive understanding, we evaluate the functions
$\rho_{{\rm Im}}$ {[}Eq.~(\ref{eq:RIm}){]}, ${\rm IPR}_{\max}$
{[}Eq.~(\ref{eq:IPRmax}){]}, ${\rm IPR}_{\min}$ {[}Eq.~(\ref{eq:IPRmin}){]}
and $\zeta$ {[}Eq.~(\ref{eq:zeta}){]} in a typical domain of the
parameter space $(J_{{\rm L}},U)$ for the AHM, with results presented
in Fig.~\ref{fig:AHMfig2}. In Fig.~\ref{fig:AHMfig2}(a), we realize
that the real-to-complex spectral transition of the noninteracting
model is essentially not affected by interactions, in the sense that
the bulk of eigenenergies turn into complex at any $U\neq0$ across
the transition point $J_{{\rm L}}=V=1.5$ of the $\hat{H}_{0}\equiv\hat{H}(U=0)$
{[}Eq.~(\ref{eq:H}){]}. However, this is not the case for the delocalization
transition. In Fig.~\ref{fig:AHMfig2}(c), we find that extended states
start to appear across the spectral transition at $J_{{\rm L}}=V=1.5$,
indicating the onset of delocalization transitions. Nevertheless,
the system could enter the extended phase only at a $J_{{\rm L}}$
higher than the noninteracting critical point $V=1.5$ for any $U\neq0$,
as shown by the transition boundary of the ${\rm IPR}_{\max}$ in
Fig.~\ref{fig:AHMfig2}(b). Therefore, onsite interactions could drive
the ending point of delocalization transition of the AHM towards a stronger
asymmetric hopping regime. Similar situations have been encountered
before in the last subsection, highlighting the generality of local-interaction
effects on NHQCs. Altogether, this means that the joint critical point
of the spectral and localization transitions of the noninteracting
AHM {[}Eq.~(\ref{eq:H}){]} would be split and extended into an intermediate mobility edge phase, carrying both
extended states of complex energies and localized states of real energies,
with two such examples shown in Figs.~\ref{fig:AHMfig1}(d) and \ref{fig:AHMfig1}(f).
The presence of such an intermediate phase is further affirmed by the smoking-gun
function $\zeta$ in Fig.~\ref{fig:AHMfig2}(d), which shows finite
values only in the region between $J_{{\rm L}}=V=1.5$ {[}the left boundary
of $\rho_{{\rm Im}}$ and ${\rm IPR}_{\min}${]} and the right boundary of
${\rm IPR}_{\max}$, i.e., a region sandwiched between the extended and localized phases.

\begin{figure}
\centering
\includegraphics[width=0.95\linewidth]{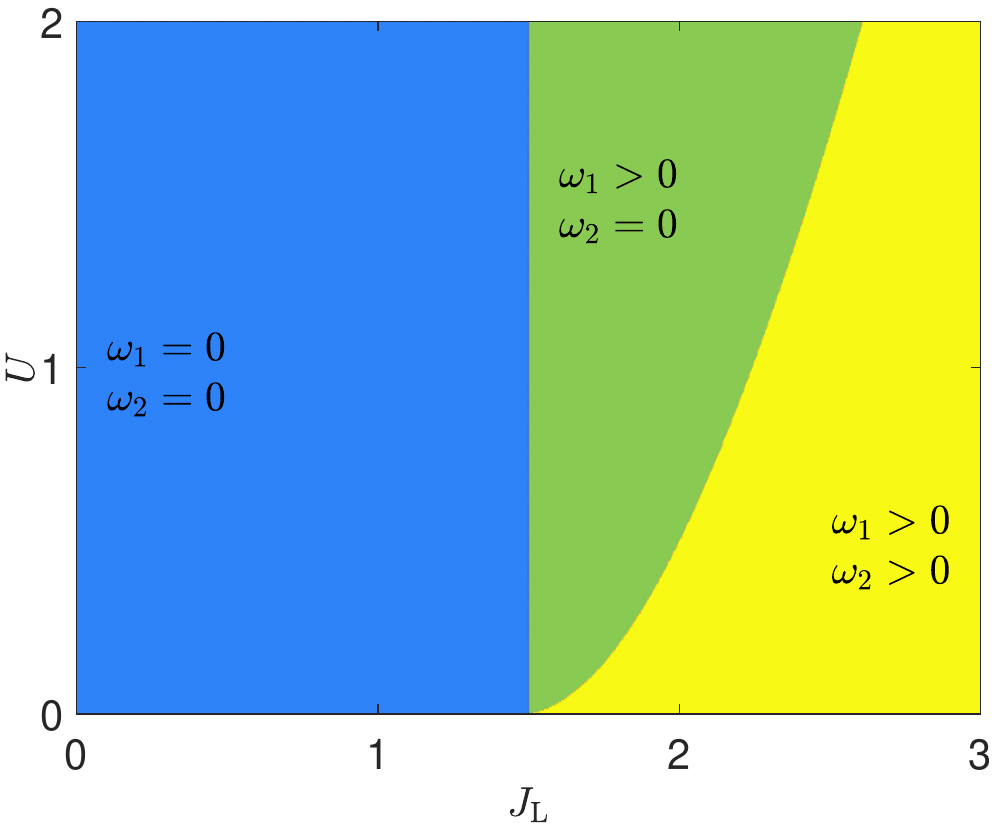}
\caption{Phase diagram of the AHM vs $(J_{\rm L},U)$ under
the PBC, characterized by the winding numbers $(\omega_1,\omega_2)$ [Eq.~(\ref{eq:w12})]. Other system parameters are $(J_{\rm R},V)=(1,1.5)$ and the length of lattice is $L=89$.}
\label{fig:AHMfig3}
\end{figure}

\begin{figure*}
	\begin{centering}
		\includegraphics[scale=0.9]{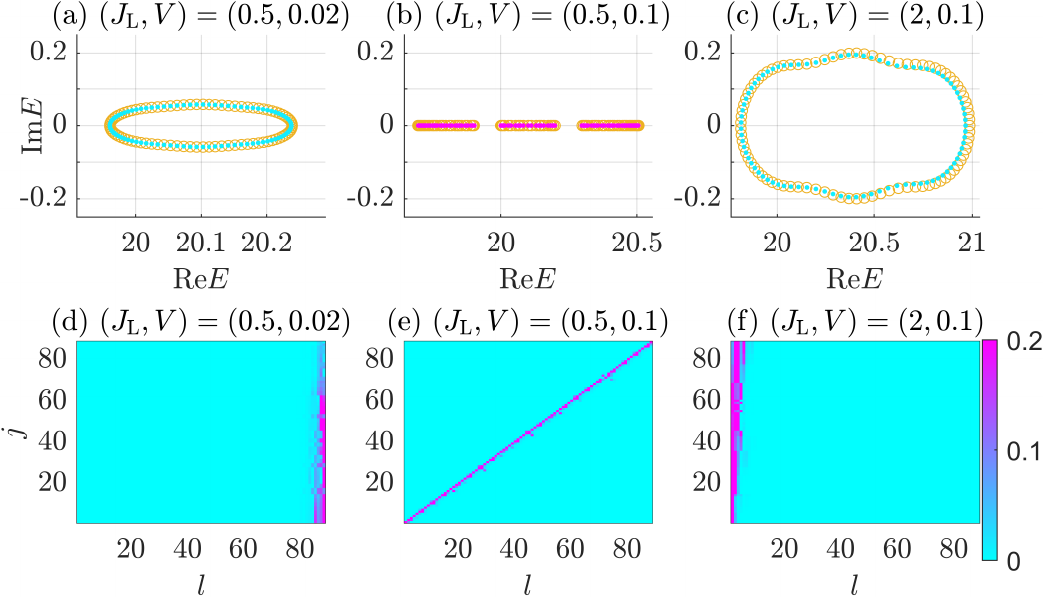}
		\par\end{centering}
	\caption{Energy spectra, IPRs and probability distributions of the bosonic
		doublons. The length of lattice $L=89$ and the system parameters $(J_{{\rm R}},U)=(1,20)$
		are taken for all panels. (a)--(f) share the same color bar. In (a)--(c),
		the dots and circles represent the spectra of $\hat{H}$ {[}Eq.~(\ref{eq:H}){]}
		in its high-energy sectors and the spectra of $2\hat{H}_{2}+U+4J_{{\rm L}}J_{{\rm R}}/U$
		{[}Eq.~(\ref{eq:H2}){]} under the PBC. The colors of dots in (a)--(c)
		give the IPRs of the corresponding states. (d)--(f) show the probability
		distributions (measured by the color bar) of all the eigenstates $|\psi_j\rangle$ of
		$\hat{H}_{2}$ under the OBC, with the horizontal and vertical axes
		representing the lattice and state indices $l$ and $j$, respectively.
		\label{fig:DBL1}}
\end{figure*}

Finally, we check the winding numbers $(\omega_{1},\omega_{2})$ in Eq.~(\ref{eq:w12})
for the AHM, with results presented in Fig.~\ref{fig:AHMfig3}. In
the localized phase (with $0\leq J_{{\rm L}}<1.5$ at any $U$), the
spectrum of AHM is real with no loops on the complex plane, and we
find $\omega_{1}=\omega_{2}=0$. In the intermediate phase, complex energy loops
start to appear in the spectrum, and we find $\omega_{1}>0$ yet $\omega_{2}=0$.
The change of $\omega_{1}$ from zero to a finite integer thus provides
the topological signature for the transition from localized to intermediate
phases. In the extended phase, we find both $\omega_{1}>0$ and $\omega_{2}>0$.
Therefore, the jump of $\omega_{2}$ from zero to a finite integer offers
the topological signal for the transition from intermediate to extended
phases. The combination of information provided by $\omega_{1}$ and $\omega_{2}$
then allows us to have a complete topological characterization of
the spectral and localization transitions in the AHM, which is also
consistent with the description based on other quantities in Fig.~\ref{fig:AHMfig2}. Note in passing that the signs of winding numbers
$(\omega_{1},\omega_{2})$ here are different from those of the UHM in the last
subsection. The underlying reason is that in the extended and intermediate
phases, the skin-localization directions of bulk states in the UHM
and AHM are opposite. Under the OBC, the eigenstates are localized at
the right edge in the extended and intermediate phases of the UHM, while they are localized at the left
edge in these phases of the AHM for our considered parameter space.

To sum up, in two typical nonreciprocal AAH quasicrystals, the cooperation
among disorder, interaction and non-Hermitian effects could indeed
create unique types of spectral, localization and topological transitions.
A key role of the onsite interaction is to dissolve the joint critical
point of single-particle spectral and localization transitions and
transform it into an intermediate mobility edge phase with both extended and localized states. In the strong-interaction regime,
this intermediate phase is expected to hold two disconnected clusters
of states. One of them contains extended two-body states, whose spectrum
looks similar to the associated noninteracting system. The other one
consists of bounded boson pairs forming doublons. We explore the spectral, localization and topological physics of these bosonic doublons in the next
section.

\section{Doublon physics}\label{sec.doublon}
In the Hamiltonian $\hat{H}$ {[}Eq.~(\ref{eq:H}){]} of AHM,
if the interaction strength $U\gg(J_{{\rm L}},J_{{\rm R}},V)$, one
may treat the noninteracting part $\hat{H}_{0}\equiv\hat{H}(U=0)$
as a perturbation to the Bose-Hubbard interaction term $\hat{H}_{{\rm int}}\equiv\sum_{l}\frac{U}{2}\hat{n}_{l}(\hat{n}_{l}-1)$.
In the Fock space of two bosons, the $\hat{H}_{{\rm int}}$ has two
sets of eigenstates, with each of them be highly degenerate. The first
set $\{|2_{l}\rangle|l=1,...,L\}$ has two bosons on the same
lattice site $l$ with the interaction energy $E=U$, which is $L$-fold
degenerate. The second set $\{|1_{l},1_{m}\rangle|l,m=1,...,L;\,l<m\}$
has two bosons on different lattice sites with the energy $E=0$,
which is $L(L-1)/2$-fold degenerate. We can also define projection
operators into these two orthogonal subspaces as
\begin{alignat}{1}
\hat{{\cal P}} & =\sum_{l}|2_{l}\rangle\langle2_{l}|,\nonumber \\
\hat{{\cal Q}} & =\sum_{l,m,l<m}\frac{1}{U}|1_{l},1_{m}\rangle\langle1_{l},1_{m}|.\label{eq:PQ}
\end{alignat}
Using the degenerate perturbation theory up to second order, we could
obtain an effective Hamiltonian in the doublon subspace $\{|2_{l}\rangle|l=1,...,L\}$,
which takes the form $\hat{H}_{2}=U\hat{{\cal P}}+\hat{{\cal P}}\hat{H}_{0}\hat{{\cal P}}+\hat{{\cal P}}\hat{H}_{0}\hat{{\cal Q}}\hat{H}_{0}\hat{{\cal P}}$ \cite{takahashi1977half,ke2017multiparticle}.
Working out the $\hat{H}_{2}$ for our AHM, we find (up to
a uniform onsite energy shift $U+4J_{{\rm L}}J_{{\rm R}}/U$ and a
global multiplication factor $2$) that
\begin{alignat}{1}
\hat{H}_{2}= & \sum_{l}\frac{1}{U}\left(J_{{\rm L}}^{2}|2_{l}\rangle\langle2_{l+1}|+J_{{\rm R}}^{2}|2_{l+1}\rangle\langle2_{l}|\right)\nonumber \\
+ & \sum_{l}2V\cos(2\pi\alpha l)|2_{l}\rangle\langle2_{l}|.\label{eq:H2}
\end{alignat}
We see that the hopping amplitudes of this doublon effective Hamiltonian
is nonreciprocal whenever $J_{{\rm L}}^{2}\neq J_{{\rm R}}^{2}$,
which means that the doublons may also subject to NHSEs under the
OBC. Under the PBC, the competition between quasiperiodic potential
and asymmetric hopping may further lead to spectral, localization
and topological transitions in the doublon subspace at
\begin{equation}
|V|=\max(J_{{\rm L}}^{2},J_{{\rm R}}^{2})/U.\label{eq:PB2}
\end{equation}
Interestingly, this condition is different from the critical point
{[}Eq.~(\ref{eq:PB}){]} of the noninteracting model. Interactions
could thus make the bosonic doublons to have different transition
thresholds than the noninteracting system.

In Fig.~\ref{fig:DBL1}, we present spectral and state profiles of
the doublons in three typical cases. First, we notice that with $U\gg(J_{{\rm L}},J_{{\rm R}},V)$,
the high-energy sectors of the spectra of $\hat{H}$ {[}Eq.~(\ref{eq:H}){]},
as denoted by the dots in Figs.~\ref{fig:DBL1}(a)--(c), could indeed
be well approximated by the spectra of the doublon effective Hamiltonian
$\hat{H}_{2}$ {[}Eq.~(\ref{eq:H2}){]}, which are given by the circles
in the corresponding figures. Second, the spectra of the doublon sectors
could either be real or complex. Specially, when $|V|$ is smaller
then $J_{{\rm L}}^{2}/U$ or $J_{{\rm R}}^{2}/U$, the spectrum is
complex and form loops on the complex plane, as shown in Figs.~\ref{fig:DBL1}(a)
and \ref{fig:DBL1}(c). When $|V|$ is larger than both $J_{{\rm L}}^{2}/U$
and $J_{{\rm R}}^{2}/U$, the spectrum is real as shown in Fig.~\ref{fig:DBL1}(b).
The condition of spectral transition in Eq.~(\ref{eq:PB2}) is thus
verified. Third, in the situation with a real (complex) spectrum in
the high-energy sector, all the eigenstates of the AHM are localized
(extended). The Eq.~(\ref{eq:PB2}) is thus expected to also describe
the critical point of localization transition for doublons. Finally,
we realize that under the OBC, all the doublon eigenstates of $\hat{H}_{2}$
become skin-localized at the right (left) edge of the lattice when
$J_{{\rm L}}^{2}<J_{{\rm R}}^{2}$ ($J_{{\rm L}}^{2}>J_{{\rm R}}^{2}$)
and $|V|<\max(J_{{\rm L}}^{2},J_{{\rm R}}^{2})/U$, as shown by their
probability distributions in Fig.~\ref{fig:DBL1}(d) (Fig.~\ref{fig:DBL1}(f)).
Therefore, by tuning the relative strengths of hopping $J_{{\rm L}}$
and $J_{{\rm R}}$, one could induce NHSEs of doublons, reversing
their skin-localization directions, and achieving pumping of doublons
from one to the other edge. Meanwhile, if $|V|>\max(J_{{\rm L}}^{2},J_{{\rm R}}^{2})/U$,
the doublon states are all constituted by strongly bounded boson pairs
pinning at different lattice sites, as illustrated in Fig.~\ref{fig:DBL1}(e).
In the AHM, one could then encounter the transition of doublon states
between skin-localizations triggered by non-Hermitian effects and
Anderson-localizations induced by quasiperiodic potentials, which
is unique to interacting and disordered non-Hermitian systems.

\begin{figure}
\begin{centering}
\includegraphics[scale=0.485]{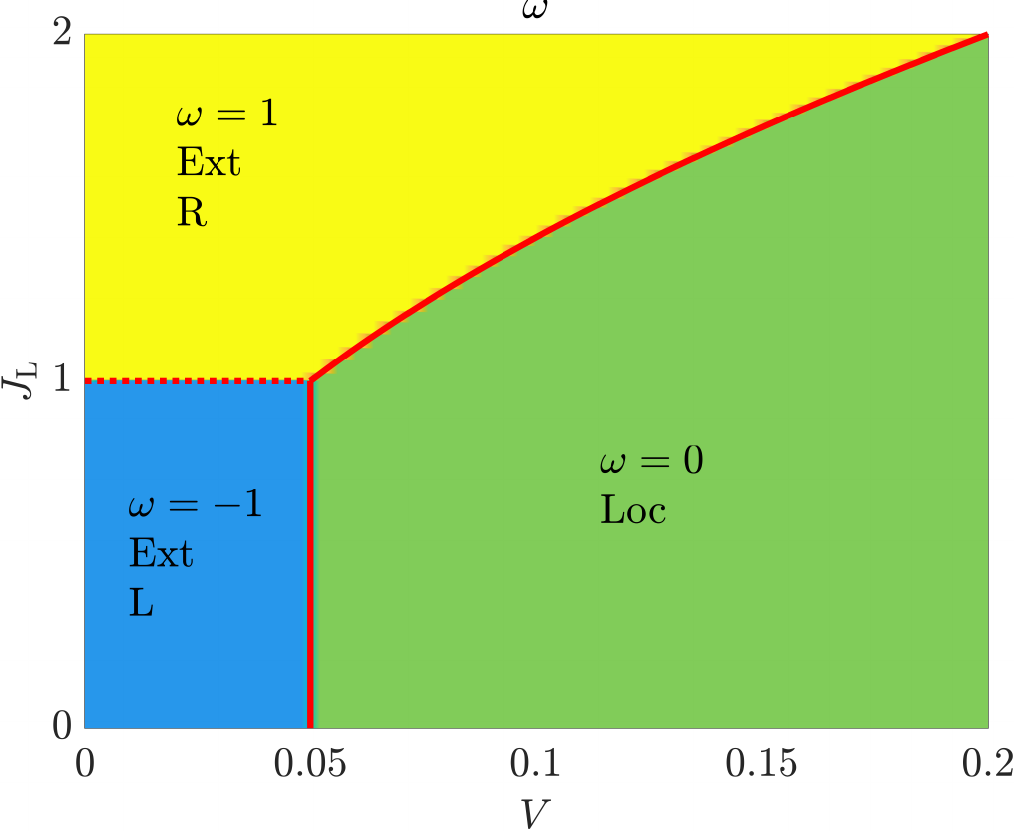}
\par\end{centering}
\caption{Phase diagram of the doublon effective Hamiltonian $\hat{H}_{2}$,
depicted by the winding number $\omega$ [Eq.~(\ref{eq:w})]. Other system parameters
are $(J_{{\rm R}},U)=(1,20)$ and the length of lattice is $L=89$.
Regions with different colors represent different phases. ``Loc''
(``Ext'') refers to the localized (extended) phase. ``L'' (``R'')
means that all the bulk states are skin-localized at the left (right)
edge of the lattice under the OBC. \label{fig:DBL2}}
\end{figure}

To complete our discussion, we show in Fig.~\ref{fig:DBL2} the phase
diagram of the doublon effective Hamiltonian $\hat{H}_{2}$, which
is obtained by computing the topological winding number $\omega$
in Eq.~(\ref{eq:w}). Via synthesizing the spectral, localization
and topological aspects, we identify three phases for doublons.
Their phase boundaries are given by the Eq.~(\ref{eq:PB2}) (the solid
line) and the condition $|J_{{\rm L}}|=|J_{{\rm R}}|$ (the dotted
line) in Fig.~\ref{fig:DBL2}. When $|V|>\max(J_{{\rm L}}^{2},J_{{\rm R}}^{2})/U$,
the doublons all form tightly bounded boson pairs with a real spectrum,
and they reside in a localized phase with the winding number $\omega=0$.
When $|V|<\max(J_{{\rm L}}^{2},J_{{\rm R}}^{2})/U$ and $|J_{{\rm L}}|<|J_{{\rm R}}|$,
the doublon states have complex eigenenergies under the PBC, and they
form an extended phase with the winding number $\omega=-1$. Under
the OBC, the doublons instead become skin-localized around the right
edge of the system. When $|V|<\max(J_{{\rm L}}^{2},J_{{\rm R}}^{2})/U$
and $|J_{{\rm L}}|>|J_{{\rm R}}|$, the doublon states also possess
complex energies under the PBC and constituting an extended phase
with the winding number $\omega=1$. Under the OBC, the doublons
are now skin-localized at the left edge of the lattice. Note in passing
that the doublon phase diagram at large $U$ in Fig.~\ref{fig:DBL2}
looks formally similar to the noninteracting single-particle phase
diagram at $U=0$ in Fig.~\ref{fig:Sketch}(d), even though the state
compositions of these two cases are rather different. Overall, the
doublon physics discussed here provides another clear evidence for the fact
that interparticle interactions could greatly enrich the phase structure
of NHQCs.

\section{Conclusion}\label{sec.conclusion}
In this study, we uncovered rich spectral and localization phenomena driven by the interplay among nonreciprocal hopping, correlated disorder and interparticle interactions in a representative class of non-Hermitian quasicrystal. With a focus on two bosons subject to Hubbard onsite interactions in AAH-type lattices, we revealed that interactions could enhance localization, specially pushing the threshold of localization transition towards stronger hopping regions. Moreover, interactions could give rise to an intermediate mobility edge phase, in which extended two-body states and bounded boson pairs are coexistent. Meanwhile, the transitions between extended, intermediate and localized phases in our system are topological in nature and characterized by a pair of integer quantized winding numbers. Finally, we analyzed the doublon states under strong interactions and unveiled their non-Hermitian skin effects, whose direction could be further controlled flexibly via tuning the asymmetric hopping parameters.

With more than two interacting bosons, there could also be interaction-induced bound states. Intermediate mobility edge phases are thus expected to emerge with the increase of disorder or interaction strengths. For example, our preliminary calculations with three interacting bosons yield an intermediate phase, in which localized triplons (three-boson bound states) coexist with extended states. They are also subject to NHSEs under the OBC. In the strong interaction regime, as the effective hopping amplitude of triplons is proportional to $J^2_{\rm L,R}/U$, they will be first localized with the increase of disorder, leading to the localization transition at weaker disorder strengths compared to the noninteracting case. Therefore, although we focus on the case of two interacting bosons in this study, the physical mechanisms underlying the interaction-induced intermediate phases and the NHSEs of multiparticle bound states we revealed are expected to hold in the cases of more than two particles.

In future work, it would be interesting to explore transitions caused by the interplay among non-Hermiticity, disorder, and many-body interactions in higher spatial dimensions. For example, the AAH model can be mapped onto the two-dimensional Hofstadter model, thereby exhibiting fractal spectra and critical eigenmodes \cite{barelli1996double,lang2012edge}. Our preliminary calculations suggest that a Hofstadter ``double-butterfly'' spectrum could appear in the non-Hermitian AAH model with two interacting bosons. The impact of interactions and non-Hermitian effects on the butterfly spectrum deserves further exploration. In non-Hermitian systems with quasiperiodic or quenched disorder, various types of entanglement phase transitions have recently been identified \cite{li2023disorder,zhou2024entanglement,li2024emergent}. The influence of interaction on these intriguing entanglement transitions forms another fascinating direction for future research.

\begin{acknowledgments} 
We acknowledge Tian Qian for helpful discussions. L.Z. is supported by the National Natural Science Foundation of China (Grants No.~12275260, No.~12047503, and No.~11905211), the Fundamental Research Funds for the Central Universities (Grant No.~202364008), and the Young Talents Project of Ocean University of China.
\end{acknowledgments}	
	
\bibliography{nhbhm.bib}

\end{document}